\documentclass[12pt]{article}
\usepackage{authblk}
\usepackage[bookmarksnumbered, colorlinks, plainpages]{hyperref}
\usepackage{amsmath, amsthm, amscd, amsfonts, amssymb, graphicx, color, booktabs,cite,url}
\newcommand{\beq}{\begin{equation}}
\newcommand{\eeq}{\end{equation}}

\numberwithin{equation}{section}
\definecolor{email}{rgb}{0.00,0.00,0.84}
\begin{document}
\setcounter{page}{1}

\title{\large \bf 12th Workshop on the CKM Unitarity Triangle\\ Santiago de Compostela, 18-22 September 2023 \\ \vspace{0.3cm}
\LARGE Working group 1 summary: $\boldsymbol{V_{ud}}$, $\boldsymbol{V_{us}}$, $\boldsymbol{V_{cd}}$, $\boldsymbol{V_{cs}}$ and semileptonic/leptonic $\boldsymbol{D}$ decays
}

\author[1]{Bipasha Chakraborty\footnote{b.chakraborty@soton.ac.uk}}
\author[2]{Alex Gilman\footnote{alex.gilman@physics.ox.ac.uk}}
\author[3]{\\ Martin Hoferichter\footnote{hoferichter@itp.unibe.ch}}
\author[4]{Michal Koval\footnote{michal.koval@matfyz.cuni.cz}}
\affil[1]{University of Southampton\linebreak
University Road, SO17 IBJ, United Kingdom}
\affil[2]{University of Oxford \linebreak
Denys Wilkinson Building \linebreak Keble Road, OX1 3RH, United Kingdom}
\affil[3]{Albert Einstein Center for Fundamental Physics,\linebreak Institute for Theoretical Physics, University of Bern,\linebreak Sidlerstrasse 5, 3012 Bern, Switzerland}
\affil[4]{Institute of Particle and Nuclear Physics,\linebreak Faculty of Mathematics and Physics, Charles University,\linebreak V Holesovickach 2, 180 00 Prague 8, Czech Republic}
\date{}

\maketitle

\begin{abstract}
We summarize the program of working group 1 at the 12th Workshop on the CKM Unitarity Triangle, whose main subjects covered $V_{ud}$, $V_{us}$, and first-row unitarity as well as $V_{cd}$, $V_{cs}$, and (semi-)leptonic $D$ decays. 
\end{abstract} 

\maketitle

\section{Introduction}

While the program of the CKM 2023 conference focused on precision tests of the CKM paradigm via unitarity triangles, a separate class of precision tests only involves the moduli of the CKM matrix elements $|V_{ij}|$, since unitarity demands that 
\beq
\sum_{j=d,s,b}|V_{ij}|^2=1,\qquad \sum_{i=u,c,t}|V_{ij}|^2=1,
\eeq
for fixed $i=u,c,t$ and $j=d,s,b$, respectively. The corresponding unitarity tests thus amount to unitarity circles, referred to as ``first-row'' and ``first-column'' unitarity. The primary goal of the work presented in WG 1 constitutes improvements in the unitarity test involving $V_{ud}$, $V_{us}$, $V_{cd}$, $V_{cs}$, i.e., first/second row/column, given that the third-generation elements therein are negligible at the current level of precision. Looking at the status as quoted in Ref.~\cite{ParticleDataGroup:2022pth}:
\begin{align}
  &\text{first row:} & |V_{ud}|^2+|V_{us}|^2+|V_{ub}|^2&=0.9984(7),\notag\\
  &\text{second row:} &|V_{cd}|^2+|V_{cs}|^2+|V_{cb}|^2&=1.001(12),\notag\\
  &\text{first column:} &|V_{ud}|^2+|V_{cd}|^2+|V_{td}|^2&=0.9971(20),\notag\\
  &\text{second column:}& |V_{us}|^2+|V_{cs}|^2+|V_{ts}|^2&=1.003(12),
 \end{align}
one observes that there are hints for a deficit in the first-row and first-column unitarity tests, motivating increased efforts both in experiment and theory to corroborate the current results and further increase the precision. Here, we provide a summary of the corresponding working group 1 program, while for some presentations, separate write-ups are available~\cite{Hoferichter:2024qsp,Wang:2024fpp,Passeri:2022lab,Schwartz:2024trf,Kellermann:2024zfy}. 

\begin{figure}[t]
	\centering
	\includegraphics[width=0.7\linewidth,clip]{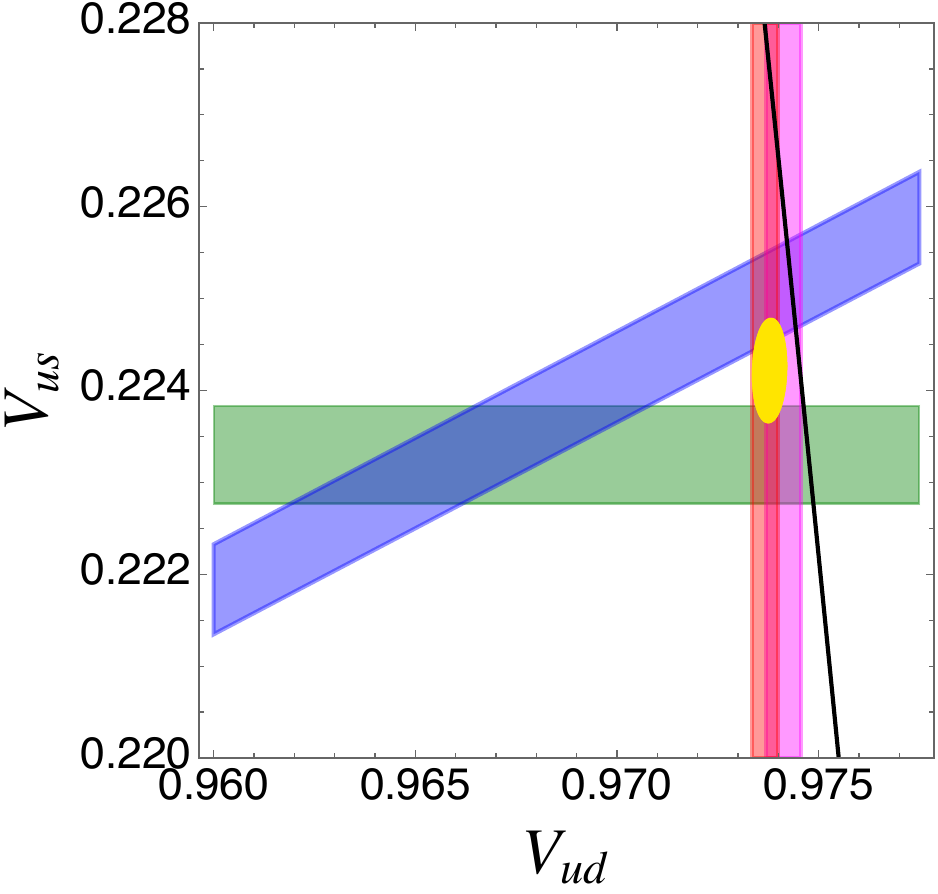}
	\caption{Current status of first-row CKM unitarity, figure taken from Ref.~\cite{Cirigliano:2022yyo}. The (green) horizontal band represents $V_{us}$ from $K_{\ell 3}$ decays, the (blue) diagonal band $V_{us}/V_{ud}$ from $K_{\ell 2}/\pi_{\ell 2}$ decays, and the vertical bands to $V_{ud}$ from superallowed $\beta$ decays (left, red) as well as neutron decay (right, violet), respectively. The unitarity circle is displayed by the black solid line, the global fit by the yellow ellipse.}
	\label{fig:CKM}
\end{figure}

\section{\texorpdfstring{$\boldsymbol{V_{ud}}$}{}, \texorpdfstring{$\boldsymbol{V_{us}}$}{}, and first-row unitarity}

The current status of the first-row unitarity test is shown in Fig.~\ref{fig:CKM}, including constraints from $K_{\ell 3}=K\to \pi\ell\nu_\ell$ decays, $\pi_{\ell 2}/K_{\ell 2}$, and $\beta$ decays. Tensions arise both between the global fit and with unitarity, but each pair of constraints by itself leads to a unitarity deficit $\Delta_\text{CKM}\equiv|V_{ud}|^2+|V_{us}|^2-1<0$~\cite{Cirigliano:2022yyo}
\begin{align}
     \Delta_\text{CKM}^{K_{\ell 2}\text{--}K_{\ell 3}} &= -0.016(6)\, [2.6\sigma],\notag\\
     \Delta_\text{CKM}^{K_{\ell 2}\text{--}\beta} &= -0.0010(6)\, [1.7\sigma],\notag\\
     \Delta_\text{CKM}^{K_{\ell 3}\text{--}\beta} &= -0.0018(6)\, [3.1\sigma],\notag\\
     \Delta_\text{CKM}^\text{global} &= -0.0018(6)\, [2.8\sigma],
     \end{align}
suggesting a tension even just within the kaon sector. Accordingly, it is clear that both $V_{ud}$ and $V_{us}$ need to be improved. 

\begin{table}[t]
 \centering
 \scalebox{0.906}{
 \begin{tabular}{lrr}\toprule
  & $+$ & $-$\\\midrule
  Superallowed $\beta$ decays & many isotopes to average & nuclear uncertainties\\
  Neutron decay &no nuclear uncertainties & need high precision for $\tau_n$ and  $g_A$\\
  Pion $\beta$ decay & theoretically pristine & experimentally challenging
  \\\bottomrule
 \end{tabular}}
\caption{Sources for $V_{ud}$ determinations, indicating some advantages and disadvantages of each method.}
\label{tab:Vud}
\end{table}

\subsection{\texorpdfstring{$\boldsymbol{V_{ud}}$}{}}

Sources for $V_{ud}$ determinations are summarized in Table~\ref{tab:Vud}, together with their respective advantages and disadvantages. In the following, we discuss recent developments for each method in more detail. 

For superallowed $\beta$ decays, one argument for the reliability of nuclear corrections brought forward in Ref.~\cite{Hardy:2020qwl} is that once all corrections are applied, one observes good consistency among a large number of nuclear transitions. To test the nuclear models that were employed traditionally, a new method to compute nuclear-structure corrections using dispersion relations was developed~\cite{Seng:2022cnq,Gorchtein:2023naa}, which was recently applied to the $^{10}\text{C}\to{}^{10}\text{B}$ transition in combination with modern ab-initio nuclear-structure methods~\cite{Gennari:2024sbn}. Moreover, measurements of nuclear charge radii could help constrain nuclear-structure calculations~\cite{Seng:2022epj}, and such measurements are possible, e.g., at PSI, FRIB, and ISOLDE.
 
A complementary approach using effective field theory (EFT) is being developed to establish a framework that allows one to explicitly take advantage of the separation of scales, resum large logarithms, and systematically improve calculations by including subleading orders in the various EFT expansions. The application to neutron decay~\cite{Cirigliano:2023fnz} was discussed in detail,  demonstrating how the different scales ranging from the mass of the $W$-boson down to the electron mass and small ${\mathcal Q}$ value of the decay have to be evolved and at which places the non-perturbative input enters. Recently, this EFT approach was extended to nuclear decays, including a first ab-initio application to $^{14}\text{O}\to{}^{14}\text{N}$~\cite{Cirigliano:2024msg,Cirigliano:2024rfk}.

For neutron decay, the uncertainty in the resulting $V_{ud}$ value is dominated by the experimental input, since both for the neutron lifetime $\tau_n$ and the asymmetry parameter $\lambda=g_A/g_V$ very high precision is required. For $\tau_n$, the precision necessary for a determination competitive with superallowed $\beta$ decays is already within reach~\cite{UCNt:2021pcg}, but $\lambda$ needs to be further improved, and, crucially, the tension between PERKEO III~\cite{Markisch:2018ndu} and aSPECT~\cite{Beck:2019xye} must be resolved. At the future PERC experiment, a precision of $\Delta\lambda/\lambda\simeq 10^{-4}$ is anticipated, which, combined with ongoing improvements in $\tau_n$ measurements, would allow for an independent, competitive $V_{ud}$ determination free of nuclear uncertainties. 

Third, $V_{ud}$ can be determined from pion $\beta$ decay, $\pi^+\to \pi^0e^+\bar\nu_e$, but its measurement at the required level of precision is extremely challenging.  
PIONEER is a next-generation rare pion decay experiment at PSI~\cite{PIONEER:2022yag} that will pursue such a measurement in Phases II+III of its physics program, while concentrating on lepton flavor universality in the ratio $R_{e/\mu}=\Gamma[\pi^+\to e^+\nu_e(\gamma)]/\Gamma[\pi^+\to \mu^+\nu_\mu(\gamma)]$ in Phase I (which could be related to CKM unitarity~\cite{Crivellin:2020lzu,Crivellin:2021njn}). For both measurements, improvements by a factor $10$ are possible before theory uncertainties become relevant, presenting unique opportunities to explore uncharted parameter space.

\subsection{\texorpdfstring{$\boldsymbol{V_{us}}$}{}}

For $V_{us}$, the results of an updated global fit to kaon decays were presented, using the methodology from Refs.~\cite{FlaviaNetWorkingGrouponKaonDecays:2010lot,Moulson:2017ive}, and leading to the situation depicted in Fig.~\ref{fig:CKM}, see Ref.~\cite{Cirigliano:2022yyo}. While new data for $K_S\to\pi e\nu$ have become available~\cite{KLOE-2:2022dot}  
\begin{equation}
\text{Br}[K_S\to\pi e\nu]=7.153(37)_\text{stat}(44)_\text{syst}\times 10^{-4},  
\end{equation}
the tension between $K_{\ell 2}$ and $K_{\ell 3}$ determinations remains, motivating a new measurement of $K_{\mu 3}/K_{\mu 2}$ at NA62 or elsewhere~\cite{Anzivino:2023bhp} to either resolve the tension or corroborate the need for a BSM explanation.  

The updated fit to $K_{\ell 2}$ and $K_{\ell 3}$ data also includes improvements in critical radiative corrections. For $K_{\ell 3}$ decays these were evaluated in a dispersive framework~\cite{Seng:2021boy,Seng:2021wcf,Seng:2022wcw}, in agreement with earlier calculations in chiral perturbation theory (ChPT)~\cite{Cirigliano:2001mk,Cirigliano:2008wn}, but reducing the uncertainty due to low-energy constants, mainly by evaluating Born-term contributions in terms of known form factors and using input from lattice QCD. For $K_{\ell 2}$ decays, the radiative corrections from ChPT~\cite{Knecht:1999ag,Cirigliano:2011tm} have been crosschecked by direct lattice-QCD calculations~\cite{DiCarlo:2019thl,Boyle:2022lsi}, and further improvements are being developed~\cite{DiCarlo:2024lue}. In the context of lattice-QCD calculations for rare kaon decays, also the prospects for $\Sigma^+\to p\ell^+\ell^-$ were discussed~\cite{Erben:2022tdu}.

\section{\texorpdfstring{$\boldsymbol{V_{cd}}$}{}, \texorpdfstring{$\boldsymbol{V_{cs}}$}{}, and (semi-)leptonic \texorpdfstring{$\boldsymbol{D}$}{} decays}

Precision on the magnitudes $|V_{cd}|$ and $|V_{cs}|$ is driven by the study of purely leptonic and semileptonic decays of charmed hadrons. The magnitude of the CKM elements can be determined from the measured decay rates with external measurements of the charmed hadron lifetime and an estimation of QCD decay constants or form factors. 

\begin{figure}[t]
    \centering
    \includegraphics[width=12cm]{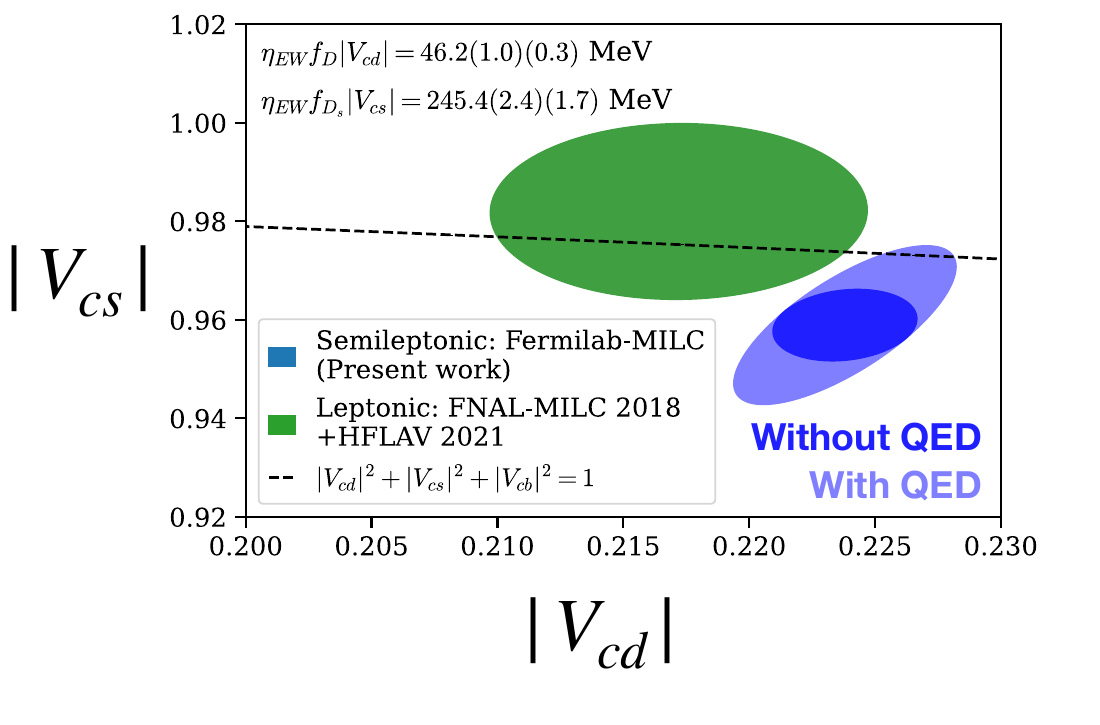}
    \caption{Updated Fermilab-MILC determination of $|V_{cs}|$ compared to the 2021 HFLAV average results~\cite{HFLAV:2022esi} from leptonic charm decays. Figure taken from Ref.~\cite{FermilabLattice:2022gku}.}
    \label{fig:Vcs}
\end{figure}

\subsection{Updated precision on \texorpdfstring{$\boldsymbol{|V_{cs}|}$}{}}
Significantly improved precision has been achieved on $|V_{cs}|$ since the last iteration of the CKM conference due to new work from experimental collaborations and new lattice determinations of QCD form factors. In the two intervening years, the BESIII collaboration published updated measurements of $D_s^+\to \mu^+ \nu_\mu$~\cite{BESIIIDsMuNu} and $D_s^+\to \tau^+ \nu_\tau$ with multiple $\tau^+$ final states~\cite{BESIIIDstauenu,BESIIIDstaupinu,BESIIIDstaurhonu}. In combination, these updated measurements determine $|V_{cs}|=0.9774(56)(72)$ with inputs of particle masses~\cite{ParticleDataGroup:2022pth}, the $D_s^+$ lifetime~\cite{ParticleDataGroup:2022pth}, and the $D_s^+$ decay constant~\cite{FlavourLatticeAveragingGroupFLAG:2021npn}. New calculations of $D$ to pseudoscalar form factors from both the Fermilab-MILC~\cite{FermilabLattice:2022gku} and HPQCD collaborations~\cite{Chakraborty:2021qav} provide significantly improved determinations of $|V_{cs}|$ from the BESIII measurements of $D^0\to K^-\ell^+ \nu_\ell $~\cite{BESIIIKenu,BESIIIKmunu}. The impacts of the new Fermilab-MILC calculation are shown in Fig.~\ref{fig:Vcs}, compared to an average determination from leptonic charm decays from HFLAV 2021~\cite{HFLAV:2022esi}, which does not include the recent BESIII results. 

The Belle II collaboration measured the lifetimes of multiple charmed mesons with significantly improved precision~\cite{Belle-II:2021cxx,Belle-II:2023eii,Belle-II:2022ggx,Belle-II:2022plj}. The updated measurements of the $D^0$, $D^+$, $D_s^+$, and $\Lambda_c^+$ lifetimes reduce uncertainties in the determination of $|V_{cd}|$ and $|V_{cs}|$.  Prospects for competitive measurements of leptonic and semileptonic charm decays at Belle II in future datasets were also presented.

\begin{figure}[t]
    \centering
    \includegraphics[width=10cm]{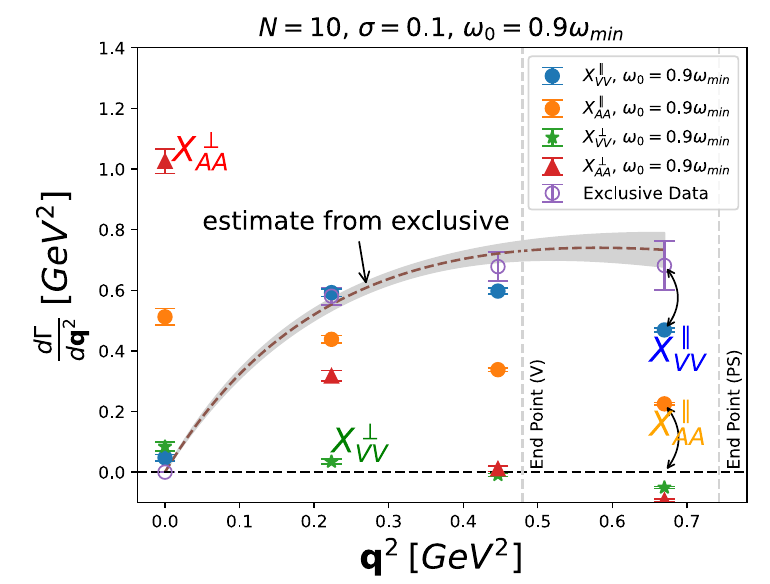}
    \caption{Different kinematical contributions to the inclusive $D_s^+\to X_s \ell^+\nu_e$ decay rate compared to exclusive lattice data~\cite{Kellermann:2022mms}.  }
    \label{fig:InclusiveLattice}
\end{figure}

\subsection{Progress on inclusive semileptonic charm decays}
A joint session with working group 2 was held on inclusive semileptonic decays of heavy hadrons. New measurements of the inclusive $D_s^+\to X e^+\nu_e$ and $\Lambda_c^+\to Xe^+\nu_e$ branching fractions and lepton momentum distributions were reported by the BESIII collaboration, both improving the precision on observables by roughly a factor of three~\cite{DsIncSL,LcIncSl}. These measurements,  in comparison to the sum of the known exclusive branching fractions~\cite{ParticleDataGroup:2022pth}, can be used to infer the existence of unobserved semileptonic decay channels. In the $D_s^+$ case, the difference between the measured inclusive branching fraction and the sum of exclusive branching fractions is $-0.04(13)(20)\%$, providing no evidence for unobserved $D_s^+$ semileptonic modes within the measured precision. However, a similar comparison for the $\Lambda_c^+$ case 
indicates the unobserved semileptonic $\Lambda_c^+$ decays~\cite{BESIII:2022qaf}.

Due to theoretical challenges, $|V_{cd}|$ and $|V_{cs}|$ have solely been determined through exclusive semileptonic decays in contrast to the beauty sector.  Recent progress in both lattice QCD and analytic theory presents exciting prospects for the extraction of $|V_{cd}|$ and $|V_{cs}|$ from inclusive semileptonic charm decays. Significant progress has been made in the study of inclusive semileptonic decays on the lattice, demonstrated in the study of $D_s^+\to X_s \ell^+\nu_e$ \cite{Kellermann:2022mms}. The determinations of the different kinematical contributions to the inclusive $D_s^+\to X_s \ell^+\nu_e$ decay rate as determined by Ref.~\cite{Kellermann:2022mms} is shown in Fig.~\ref{fig:InclusiveLattice}. Progress in the heavy quark effective theory treatment of inclusive charm decays were also discussed \cite{Fael:2019umf}, similar to that employed in beauty decays as, e.g., in Ref.~\cite{Bernlochner:2022ucr}. This progress from the theory and the lattice indicates exciting prospects for the study of inclusive semileptonic charm decays and encourages detailed experimental analysis of these processes.

\section{Conclusions}

Intriguing tensions of almost $3\sigma$ persist in the first-row test of CKM unitarity, both of the global fit with the unitarity circle and among different constraints on $V_{ud}$ and $V_{us}$. Several ways to corroborate or refute these tensions were discussed, for $V_{ud}$ ranging from improved theory for radiative corrections to superallowed $\beta$ decays and neutron decay, over experimental prospects for competitive measurements of neutron lifetime and decay asymmetry, to a long-term vision for a precision measurement of pion $\beta$ decay. For $V_{us}$ also improved radiative corrections for both $K_{\ell 2}$ and $K_{\ell 3}$ decays have become available, as well as new data on $K_S\to \pi e \nu$, leading to a situation that urgently calls for a new measurement of $K_{\mu 2}/K_{\mu 3}$ at NA62 or elsewhere.     

The total uncertainty on determinations of $|V_{cs}|$ from direct measurements was reduced by a factor of two, achieving sub-percent precision since the last iteration of the CKM conference. Promising prospects for further precision on $|V_{cs}|$ and $|V_{cd}|$ were presented, including the growing data sample at Belle II, large new datasets collected near the $\psi(3770)$ resonance by BESIII, and work from the lattice and analytic theory communities in analyzing charm decays. These could lead to competitive tests of unitarity from the second row and the first and second columns of the CKM matrix.

\section*{Acknowledgments}
We would like to thank the local organizers for putting together a great conference, and all the speakers in our working group for their contributions. 
 Financial support by the SNSF (Project No.\ PCEFP2\_181117), the Czech Science Foundation (Project No.\ 23-06770S), Charles University (Project PRIMUS 23/SCI/025), Leverhulme Trust (ECF-2019-223 G100820), STFC (Grant no.\ ST/X000583/1), STFC (Grant no.\ ST/W006251/1), and EPSRC (Grant no.\ EP/W032635/1) is gratefully acknowledged.

\bibliographystyle{JHEP}
\bibliography{References}

\providecommand{\href}[2]{#2}\begingroup\raggedright\begin{thebibliography}{10}

\bibitem{ParticleDataGroup:2022pth}
{\scshape Particle Data Group} collaboration, \emph{{Review of Particle
  Physics}}, \href{https://doi.org/10.1093/ptep/ptac097}{\emph{PTEP} {\bfseries
  2022} (2022) 083C01}.

\bibitem{Hoferichter:2024qsp}
M.~Hoferichter, \emph{{Prospects for PIONEER}},  in \emph{{12th International
  Workshop on the CKM Unitarity Triangle}}, 2024
  [\href{https://arxiv.org/abs/2403.18889}{{\ttfamily 2403.18889}}].

\bibitem{Wang:2024fpp}
{\scshape BESIII} collaboration, \emph{{Experimental status of $|V_{cd}|$ and
  $|V_{cs}|$}},  in \emph{{12th International Workshop on the CKM Unitarity
  Triangle}}, 2024 [\href{https://arxiv.org/abs/2404.00592}{{\ttfamily
  2404.00592}}].

\bibitem{Passeri:2022lab}
{\scshape KLOE-2} collaboration, \emph{{A new $K_S \to \pi e \nu$ branching
  fraction measurement from KLOE-2}},  in \emph{{12th International Workshop on
  the CKM Unitarity Triangle}}, 2024
  [\href{https://arxiv.org/abs/2404.01826}{{\ttfamily 2404.01826}}].

\bibitem{Schwartz:2024trf}
A.J.~Schwartz, \emph{{Charm lifetime measurements at Belle II}},  in
  \emph{{12th International Workshop on the CKM Unitarity Triangle}}, 2024
  [\href{https://arxiv.org/abs/2405.10394}{{\ttfamily 2405.10394}}].

\bibitem{Kellermann:2024zfy}
R.~Kellermann, A.~Barone, S.~Hashimoto, A.~J\"uttner and T.~Kaneko,
  \emph{{Updates on inclusive charmed and bottomed meson decays from the
  lattice}},  in \emph{{12th International Workshop on the CKM Unitarity
  Triangle}}, 2024 [\href{https://arxiv.org/abs/2405.06152}{{\ttfamily
  2405.06152}}].

\bibitem{Cirigliano:2022yyo}
V.~Cirigliano, A.~Crivellin, M.~Hoferichter and M.~Moulson, \emph{{Scrutinizing
  CKM unitarity with a new measurement of the $K_{\mu3}/K_{\mu2}$ branching
  fraction}}, \href{https://doi.org/10.1016/j.physletb.2023.137748}{\emph{Phys.
  Lett. B} {\bfseries 838} (2023) 137748}
  [\href{https://arxiv.org/abs/2208.11707}{{\ttfamily 2208.11707}}].

\bibitem{Hardy:2020qwl}
J.C.~Hardy and I.S.~Towner, \emph{{Superallowed $0^+ \to 0^+$ nuclear $\beta$
  decays: 2020 critical survey, with implications for V$_{ud}$ and CKM
  unitarity}}, \href{https://doi.org/10.1103/PhysRevC.102.045501}{\emph{Phys.
  Rev. C} {\bfseries 102} (2020) 045501}.

\bibitem{Seng:2022cnq}
C.-Y.~Seng and M.~Gorchtein, \emph{{Dispersive formalism for the nuclear
  structure correction $\delta_\text{NS}$ to the $\beta$ decay rate}},
  \href{https://doi.org/10.1103/PhysRevC.107.035503}{\emph{Phys. Rev. C}
  {\bfseries 107} (2023) 035503}
  [\href{https://arxiv.org/abs/2211.10214}{{\ttfamily 2211.10214}}].

\bibitem{Gorchtein:2023naa}
M.~Gorchtein and C.Y.~Seng, \emph{{Superallowed nuclear beta decays and
  precision tests of the Standard Model}},
  \href{https://doi.org/10.1146/annurev-nucl-102622-020726}{\emph{Ann. Rev.
  Nucl. Part. Sci.} {\bfseries 74} (2024) 23}
  [\href{https://arxiv.org/abs/2311.00044}{{\ttfamily 2311.00044}}].

\bibitem{Gennari:2024sbn}
M.~Gennari, M.~Drissi, M.~Gorchtein, P.~Navratil and C.-Y.~Seng, \emph{{An
  $\textit{ab initio}$ recipe for taming nuclear-structure dependence of $
  V_{ud} $: the $ {}^{10}\mathrm{C} \rightarrow {}^{10}\mathrm{B} $
  superallowed transition}},
  \href{https://arxiv.org/abs/2405.19281}{{\ttfamily 2405.19281}}.

\bibitem{Seng:2022epj}
C.-Y.~Seng and M.~Gorchtein, \emph{{Electroweak nuclear radii constrain the
  isospin breaking correction to $V_{ud}$}},
  \href{https://doi.org/10.1016/j.physletb.2022.137654}{\emph{Phys. Lett. B}
  {\bfseries 838} (2023) 137654}
  [\href{https://arxiv.org/abs/2208.03037}{{\ttfamily 2208.03037}}].

\bibitem{Cirigliano:2023fnz}
V.~Cirigliano, W.~Dekens, E.~Mereghetti and O.~Tomalak, \emph{{Effective field
  theory for radiative corrections to charged-current processes: Vector
  coupling}}, \href{https://doi.org/10.1103/PhysRevD.108.053003}{\emph{Phys.
  Rev. D} {\bfseries 108} (2023) 053003}
  [\href{https://arxiv.org/abs/2306.03138}{{\ttfamily 2306.03138}}].

\bibitem{Cirigliano:2024msg}
V.~Cirigliano, W.~Dekens, J.~de~Vries, S.~Gandolfi, M.~Hoferichter and
  E.~Mereghetti, \emph{{Ab-initio electroweak corrections to superallowed
  $\beta$ decays and their impact on $V_{ud}$}},
  \href{https://arxiv.org/abs/2405.18464}{{\ttfamily 2405.18464}}.

\bibitem{Cirigliano:2024rfk}
V.~Cirigliano, W.~Dekens, J.~de~Vries, S.~Gandolfi, M.~Hoferichter and
  E.~Mereghetti, \emph{{Radiative corrections to superallowed $\beta$ decays in
  effective field theory}},  \href{https://arxiv.org/abs/2405.18469}{{\ttfamily
  2405.18469}}.

\bibitem{UCNt:2021pcg}
{\scshape UCN$\tau$} collaboration, \emph{{Improved neutron lifetime
  measurement with UCN$\tau$}},
  \href{https://doi.org/10.1103/PhysRevLett.127.162501}{\emph{Phys. Rev. Lett.}
  {\bfseries 127} (2021) 162501}
  [\href{https://arxiv.org/abs/2106.10375}{{\ttfamily 2106.10375}}].

\bibitem{Markisch:2018ndu}
B.~M\"arkisch et~al., \emph{{Measurement of the Weak Axial-Vector Coupling
  Constant in the Decay of Free Neutrons Using a Pulsed Cold Neutron Beam}},
  \href{https://doi.org/10.1103/PhysRevLett.122.242501}{\emph{Phys. Rev. Lett.}
  {\bfseries 122} (2019) 242501}
  [\href{https://arxiv.org/abs/1812.04666}{{\ttfamily 1812.04666}}].

\bibitem{Beck:2019xye}
M.~Beck et~al., \emph{{Improved determination of the $\beta$-$\overline{\nu}_e$
  angular correlation coefficient $a$ in free neutron decay with the $aSPECT$
  spectrometer}},
  \href{https://doi.org/10.1103/PhysRevC.101.055506}{\emph{Phys. Rev. C}
  {\bfseries 101} (2020) 055506}
  [\href{https://arxiv.org/abs/1908.04785}{{\ttfamily 1908.04785}}].

\bibitem{PIONEER:2022yag}
{\scshape PIONEER} collaboration, \emph{{PIONEER: Studies of Rare Pion
  Decays}},  \href{https://arxiv.org/abs/2203.01981}{{\ttfamily 2203.01981}}.

\bibitem{Crivellin:2020lzu}
A.~Crivellin and M.~Hoferichter, \emph{{$\beta$ Decays as Sensitive Probes of
  Lepton Flavor Universality}},
  \href{https://doi.org/10.1103/PhysRevLett.125.111801}{\emph{Phys. Rev. Lett.}
  {\bfseries 125} (2020) 111801}
  [\href{https://arxiv.org/abs/2002.07184}{{\ttfamily 2002.07184}}].

\bibitem{Crivellin:2021njn}
A.~Crivellin, M.~Hoferichter and C.A.~Manzari, \emph{{Fermi Constant from Muon
  Decay Versus Electroweak Fits and Cabibbo-Kobayashi-Maskawa Unitarity}},
  \href{https://doi.org/10.1103/PhysRevLett.127.071801}{\emph{Phys. Rev. Lett.}
  {\bfseries 127} (2021) 071801}
  [\href{https://arxiv.org/abs/2102.02825}{{\ttfamily 2102.02825}}].

\bibitem{FlaviaNetWorkingGrouponKaonDecays:2010lot}
{\scshape FlaviaNet Working Group on Kaon Decays} collaboration, \emph{{An
  Evaluation of $|V_{us}|$ and precise tests of the Standard Model from world
  data on leptonic and semileptonic kaon decays}},
  \href{https://doi.org/10.1140/epjc/s10052-010-1406-3}{\emph{Eur. Phys. J. C}
  {\bfseries 69} (2010) 399} [\href{https://arxiv.org/abs/1005.2323}{{\ttfamily
  1005.2323}}].

\bibitem{Moulson:2017ive}
M.~Moulson, \emph{{Experimental determination of $V_{us}$ from kaon decays}},
  \href{https://doi.org/10.22323/1.291.0033}{\emph{PoS} {\bfseries CKM2016}
  (2017) 033} [\href{https://arxiv.org/abs/1704.04104}{{\ttfamily
  1704.04104}}].

\bibitem{KLOE-2:2022dot}
{\scshape KLOE-2} collaboration, \emph{{Measurement of the $K_S\to\pi e\nu$
  branching fraction with the KLOE experiment}},
  \href{https://doi.org/10.1007/JHEP02(2023)098}{\emph{JHEP} {\bfseries 02}
  (2023) 098} [\href{https://arxiv.org/abs/2208.04872}{{\ttfamily
  2208.04872}}].

\bibitem{Anzivino:2023bhp}
G.~Anzivino et~al., \emph{{Workshop summary: Kaons@CERN 2023}},
  \href{https://doi.org/10.1140/epjc/s10052-024-12565-4}{\emph{Eur. Phys. J. C}
  {\bfseries 84} (2024) 377}
  [\href{https://arxiv.org/abs/2311.02923}{{\ttfamily 2311.02923}}].

\bibitem{Seng:2021boy}
C.-Y.~Seng, D.~Galviz, M.~Gorchtein and U.-G.~Mei\ss{}ner,
  \emph{{High-precision determination of the $K_{e3}$ radiative corrections}},
  \href{https://doi.org/10.1016/j.physletb.2021.136522}{\emph{Phys. Lett. B}
  {\bfseries 820} (2021) 136522}
  [\href{https://arxiv.org/abs/2103.00975}{{\ttfamily 2103.00975}}].

\bibitem{Seng:2021wcf}
C.-Y.~Seng, D.~Galviz, M.~Gorchtein and U.-G.~Mei\ss{}ner, \emph{{Improved
  K$_{e3}$ radiative corrections sharpen the $K_{\mu2}$--$K_{l3}$
  discrepancy}}, \href{https://doi.org/10.1007/JHEP11(2021)172}{\emph{JHEP}
  {\bfseries 11} (2021) 172}
  [\href{https://arxiv.org/abs/2103.04843}{{\ttfamily 2103.04843}}].

\bibitem{Seng:2022wcw}
C.-Y.~Seng, D.~Galviz, M.~Gorchtein and U.-G.~Mei\ss{}ner, \emph{{Complete
  theory of radiative corrections to $K_{\ell 3}$ decays and the $V_{us}$
  update}}, \href{https://doi.org/10.1007/JHEP07(2022)071}{\emph{JHEP}
  {\bfseries 07} (2022) 071}
  [\href{https://arxiv.org/abs/2203.05217}{{\ttfamily 2203.05217}}].

\bibitem{Cirigliano:2001mk}
V.~Cirigliano, M.~Knecht, H.~Neufeld, H.~Rupertsberger and P.~Talavera,
  \emph{{Radiative corrections to $K_{l3}$ decays}},
  \href{https://doi.org/10.1007/s100520100825}{\emph{Eur. Phys. J. C}
  {\bfseries 23} (2002) 121}
  [\href{https://arxiv.org/abs/hep-ph/0110153}{{\ttfamily hep-ph/0110153}}].

\bibitem{Cirigliano:2008wn}
V.~Cirigliano, M.~Giannotti and H.~Neufeld, \emph{{Electromagnetic effects in
  $K_{l3}$ decays}},
  \href{https://doi.org/10.1088/1126-6708/2008/11/006}{\emph{JHEP} {\bfseries
  11} (2008) 006} [\href{https://arxiv.org/abs/0807.4507}{{\ttfamily
  0807.4507}}].

\bibitem{Knecht:1999ag}
M.~Knecht, H.~Neufeld, H.~Rupertsberger and P.~Talavera, \emph{{Chiral
  perturbation theory with virtual photons and leptons}},
  \href{https://doi.org/10.1007/s100529900265}{\emph{Eur. Phys. J. C}
  {\bfseries 12} (2000) 469}
  [\href{https://arxiv.org/abs/hep-ph/9909284}{{\ttfamily hep-ph/9909284}}].

\bibitem{Cirigliano:2011tm}
V.~Cirigliano and H.~Neufeld, \emph{{A note on isospin violation in
  $P_{l2(\gamma)}$ decays}},
  \href{https://doi.org/10.1016/j.physletb.2011.04.038}{\emph{Phys. Lett. B}
  {\bfseries 700} (2011) 7} [\href{https://arxiv.org/abs/1102.0563}{{\ttfamily
  1102.0563}}].

\bibitem{DiCarlo:2019thl}
M.~Di~Carlo et~al., \emph{{Light-meson leptonic decay rates in lattice
  QCD+QED}}, \href{https://doi.org/10.1103/PhysRevD.100.034514}{\emph{Phys.
  Rev. D} {\bfseries 100} (2019) 034514}
  [\href{https://arxiv.org/abs/1904.08731}{{\ttfamily 1904.08731}}].

\bibitem{Boyle:2022lsi}
P.~Boyle et~al., \emph{{Isospin-breaking corrections to light-meson leptonic
  decays from lattice simulations at physical quark masses}},
  \href{https://doi.org/10.1007/JHEP02(2023)242}{\emph{JHEP} {\bfseries 02}
  (2023) 242} [\href{https://arxiv.org/abs/2211.12865}{{\ttfamily
  2211.12865}}].

\bibitem{DiCarlo:2024lue}
M.~Di~Carlo, \emph{{Isospin-breaking corrections to weak decays: the current
  status and a new infrared improvement}},
  \href{https://doi.org/10.22323/1.453.0120}{\emph{PoS} {\bfseries LATTICE2023}
  (2024) 120} [\href{https://arxiv.org/abs/2401.07666}{{\ttfamily
  2401.07666}}].

\bibitem{Erben:2022tdu}
F.~Erben, V.~G\"ulpers, M.T.~Hansen, R.~Hodgson and A.~Portelli,
  \emph{{Prospects for a lattice calculation of the rare decay $\Sigma^+\to
  p\ell^+\ell^-$}}, \href{https://doi.org/10.1007/JHEP04(2023)108}{\emph{JHEP}
  {\bfseries 04} (2023) 108}
  [\href{https://arxiv.org/abs/2209.15460}{{\ttfamily 2209.15460}}].

\bibitem{HFLAV:2022esi}
{\scshape HFLAV} collaboration, \emph{{Averages of b-hadron, c-hadron, and
  \ensuremath{\tau}-lepton properties as of 2021}},
  \href{https://doi.org/10.1103/PhysRevD.107.052008}{\emph{Phys. Rev. D}
  {\bfseries 107} (2023) 052008}
  [\href{https://arxiv.org/abs/2206.07501}{{\ttfamily 2206.07501}}].

\bibitem{FermilabLattice:2022gku}
{\scshape Fermilab Lattice, MILC} collaboration, \emph{{D-meson semileptonic
  decays to pseudoscalars from four-flavor lattice QCD}},
  \href{https://doi.org/10.1103/PhysRevD.107.094516}{\emph{Phys. Rev. D}
  {\bfseries 107} (2023) 094516}
  [\href{https://arxiv.org/abs/2212.12648}{{\ttfamily 2212.12648}}].

\bibitem{BESIIIDsMuNu}
{\scshape BESIII} collaboration, \emph{{Improved measurement of the branching
  fraction of $D_s^+\to\mu^+\nu_\mu$}},
  \href{https://doi.org/10.1103/PhysRevD.108.112001}{\emph{Phys. Rev. D}
  {\bfseries 108} (2023) 112001}
  [\href{https://arxiv.org/abs/2307.14585}{{\ttfamily 2307.14585}}].

\bibitem{BESIIIDstauenu}
{\scshape BESIII} collaboration, \emph{{Measurement of the Absolute Branching
  Fraction of $D_s^+ \to \tau^+ \nu_{\tau}$ via $\tau^+ \to e^+ \nu_e
  \bar{\nu}_{\tau}$}},
  \href{https://doi.org/10.1103/PhysRevLett.127.171801}{\emph{Phys. Rev. Lett.}
  {\bfseries 127} (2021) 171801}
  [\href{https://arxiv.org/abs/2106.02218}{{\ttfamily 2106.02218}}].

\bibitem{BESIIIDstaupinu}
{\scshape BESIII} collaboration, \emph{{Measurement of the absolute branching
  fractions for purely leptonic $D_s^+$ decays}},
  \href{https://doi.org/10.1103/PhysRevD.104.052009}{\emph{Phys. Rev. D}
  {\bfseries 104} (2021) 052009}
  [\href{https://arxiv.org/abs/2102.11734}{{\ttfamily 2102.11734}}].

\bibitem{BESIIIDstaurhonu}
{\scshape BESIII} collaboration, \emph{{Measurement of the branching fraction
  of leptonic decay $D_s^+\to\tau^+\nu_\tau$ via $\tau^+\to\pi^+\pi^0\bar
  \nu_\tau$}}, \href{https://doi.org/10.1103/PhysRevD.104.032001}{\emph{Phys.
  Rev. D} {\bfseries 104} (2021) 032001}
  [\href{https://arxiv.org/abs/2105.07178}{{\ttfamily 2105.07178}}].

\bibitem{FlavourLatticeAveragingGroupFLAG:2021npn}
{\scshape Flavour Lattice Averaging Group (FLAG)} collaboration, \emph{{FLAG
  Review 2021}},
  \href{https://doi.org/10.1140/epjc/s10052-022-10536-1}{\emph{Eur. Phys. J. C}
  {\bfseries 82} (2022) 869}
  [\href{https://arxiv.org/abs/2111.09849}{{\ttfamily 2111.09849}}].

\bibitem{Chakraborty:2021qav}
{\scshape HPQCD} collaboration, \emph{{Improved $V_{cs}$ determination using
  precise lattice QCD form factors for $D\to K\ell\nu$}},
  \href{https://doi.org/10.1103/PhysRevD.104.034505}{\emph{Phys. Rev. D}
  {\bfseries 104} (2021) 034505}
  [\href{https://arxiv.org/abs/2104.09883}{{\ttfamily 2104.09883}}].

\bibitem{BESIIIKenu}
{\scshape BESIII} collaboration, \emph{{Study of Dynamics of $D^0 \to K^- e^+
  \nu_{e}$ and $D^0\to\pi^- e^+ \nu_{e}$ Decays}},
  \href{https://doi.org/10.1103/PhysRevD.92.072012}{\emph{Phys. Rev. D}
  {\bfseries 92} (2015) 072012}
  [\href{https://arxiv.org/abs/1508.07560}{{\ttfamily 1508.07560}}].

\bibitem{BESIIIKmunu}
{\scshape BESIII} collaboration, \emph{{Study of the $D^0\to K^-\mu^+\nu_\mu$
  dynamics and test of lepton flavor universality with $D^0\to
  K^-\ell^+\nu_\ell$ decays}},
  \href{https://doi.org/10.1103/PhysRevLett.122.011804}{\emph{Phys. Rev. Lett.}
  {\bfseries 122} (2019) 011804}
  [\href{https://arxiv.org/abs/1810.03127}{{\ttfamily 1810.03127}}].

\bibitem{Belle-II:2021cxx}
{\scshape Belle-II} collaboration, \emph{{Precise measurement of the $D^0$ and
  $D^+$ lifetimes at Belle II}},
  \href{https://doi.org/10.1103/PhysRevLett.127.211801}{\emph{Phys. Rev. Lett.}
  {\bfseries 127} (2021) 211801}
  [\href{https://arxiv.org/abs/2108.03216}{{\ttfamily 2108.03216}}].

\bibitem{Belle-II:2023eii}
{\scshape Belle-II} collaboration, \emph{{Precise Measurement of the $D_s^+$
  Lifetime at Belle II}},
  \href{https://doi.org/10.1103/PhysRevLett.131.171803}{\emph{Phys. Rev. Lett.}
  {\bfseries 131} (2023) 171803}
  [\href{https://arxiv.org/abs/2306.00365}{{\ttfamily 2306.00365}}].

\bibitem{Belle-II:2022ggx}
{\scshape Belle-II} collaboration, \emph{{Measurement of the $\Lambda_c^+$
  Lifetime}}, \href{https://doi.org/10.1103/PhysRevLett.130.071802}{\emph{Phys.
  Rev. Lett.} {\bfseries 130} (2023) 071802}
  [\href{https://arxiv.org/abs/2206.15227}{{\ttfamily 2206.15227}}].

\bibitem{Belle-II:2022plj}
{\scshape Belle-II} collaboration, \emph{{Measurement of the $\Omega_c^0$
  lifetime at Belle II}},
  \href{https://doi.org/10.1103/PhysRevD.107.L031103}{\emph{Phys. Rev. D}
  {\bfseries 107} (2023) L031103}
  [\href{https://arxiv.org/abs/2208.08573}{{\ttfamily 2208.08573}}].

\bibitem{Kellermann:2022mms}
R.~Kellermann, A.~Barone, S.~Hashimoto, A.~J\"uttner and T.~Kaneko,
  \emph{{Inclusive semi-leptonic decays of charmed mesons with M\"obius domain
  wall fermions}}, \href{https://doi.org/10.22323/1.430.0414}{\emph{PoS}
  {\bfseries LATTICE2022} (2023) 414}
  [\href{https://arxiv.org/abs/2211.16830}{{\ttfamily 2211.16830}}].

\bibitem{DsIncSL}
{\scshape BESIII} collaboration, \emph{{Measurement of the absolute branching
  fraction of inclusive semielectronic $D_s^+$ decays}},
  \href{https://doi.org/10.1103/PhysRevD.104.012003}{\emph{Phys. Rev. D}
  {\bfseries 104} (2021) 012003}
  [\href{https://arxiv.org/abs/2104.07311}{{\ttfamily 2104.07311}}].

\bibitem{LcIncSl}
{\scshape BESIII} collaboration, \emph{{Improved measurement of the absolute
  branching fraction of inclusive semileptonic $\Lambda_c^+$ decay}},
  \href{https://doi.org/10.1103/PhysRevD.107.052005}{\emph{Phys. Rev. D}
  {\bfseries 107} (2023) 052005}
  [\href{https://arxiv.org/abs/2212.03753}{{\ttfamily 2212.03753}}].

\bibitem{BESIII:2022qaf}
{\scshape BESIII} collaboration, \emph{{First observation of the semileptonic
  decay $\Lambda_c^+\to p K^-e^+\nu_e$}},
  \href{https://doi.org/10.1103/PhysRevD.106.112010}{\emph{Phys. Rev. D}
  {\bfseries 106} (2022) 112010}
  [\href{https://arxiv.org/abs/2207.11483}{{\ttfamily 2207.11483}}].

\bibitem{Fael:2019umf}
M.~Fael, T.~Mannel and K.K.~Vos, \emph{{The Heavy Quark Expansion for Inclusive
  Semileptonic Charm Decays Revisited}},
  \href{https://doi.org/10.1007/JHEP12(2019)067}{\emph{JHEP} {\bfseries 12}
  (2019) 067} [\href{https://arxiv.org/abs/1910.05234}{{\ttfamily
  1910.05234}}].

\bibitem{Bernlochner:2022ucr}
F.~Bernlochner, M.~Fael, K.~Olschewsky, E.~Persson, R.~van Tonder, K.K.~Vos
  et~al., \emph{{First extraction of inclusive V$_{cb}$ from q$^{2}$ moments}},
  \href{https://doi.org/10.1007/JHEP10(2022)068}{\emph{JHEP} {\bfseries 10}
  (2022) 068} [\href{https://arxiv.org/abs/2205.10274}{{\ttfamily
  2205.10274}}].

\end{thebibliography}\endgroup

\end{document}